\documentstyle[aps,multicol,eqsecnum]{revtex}

\begin{document}

\title{Gravity in the brane-world 
for two-branes model with stabilized modulus}
\author{Takahiro Tanaka$^{1,2}$ and Xavier Montes$^{1}$\\$~$}
\address{$^{1}$IFAE, Departament de Fisica, Universitat Autonoma de Barcelona,\\
08193 Bellaterra $($Barcelona$)$, Spain$;$\\
$^{2}$Department of Earth and Space Science, Graduate School of Science\\
 Osaka University, Toyonaka 560-0043, Japan.
}

\maketitle

\thispagestyle{empty}
\begin{abstract}
  We present a complete scheme to discuss linear perturbations in
  the two-branes model of the Randall and Sundrum scenario with the
  stabilization mechanism proposed by Goldberger and Wise.  
  We confirm that under the approximation of zero mode truncation 
  the induced metric on the branes reproduces that of the usual 
  4-dimensional Einstein gravity. We also present formulas to 
  evaluate the mass spectrum and the contribution to the metric 
  perturbations from all the Kaluza-Klein modes. We also conjecture 
  that the model has tachyonic modes unless the background configuration 
  for the bulk scalar field introduced to stabilize the distance between 
  the two branes is monotonic in the fifth dimension. 
\end{abstract}

\begin{multicols}{2}
\section{introduction}
Recently, there has been a growing interest in considering 
extra-dimensions in non-trivial form\cite{HW,AH,RS1,RS2}.  One
important suggestion was done by Randall and Sundrum\cite{RS1}.
They realized that if one considers 5-dimensional gravity with 
a negative cosmological constant bounded by two positive and 
negative tension branes, the induced
gravity on the negative tension brane can possess a very small
gravitational constant as compared with the energy scale introduced in
the original Lagrangian.  This fact potentially gives a solution to
the hierarchy problem of explaining the extraordinary weakness of the
gravitational coupling.

Later, the same authors pointed out that 
the model with the positive tension brane alone 
is also possible\cite{RS2}. 
In this case, the extension of the extra dimension is infinite. 
Nevertheless, the induced gravity on the brane can 
be expected to mimic Einstein gravity.  

There are many discussions about the cosmology based on these scenarios
\cite{Bin,Fla,CGRT,GS,MSM,MWBH}.  The behavior of gravity on these models
has been also investigated by many
authors\cite{GT,SMS1,SMS2,CG,CHR,EHM,Chiba}. 
In this direction, an explicit method to deal with the linear order
metric perturbations induced by the matter fields confined on the
branes was developed in the paper by Garriga and Tanaka\cite{GT}(Paper
I).  In that paper, considering the zero mode truncation
approximation, it was shown that the induced gravity on the branes
becomes of the Brans-Dicke type in the two-branes model. On the other hand,
Einstein gravity is recovered in the single-brane model.

However, the analysis of the two-branes model in Paper I was not
complete. The stabilization mechanism of the distance between the two
branes\cite{GW,DeW} by means of a bulk scalar field was not taken
into account. Moreover, the Kaluza-Klein contribution was not 
estimated for the two-branes model.  In this paper we consider a model
with stabilization mechanism, and we confirm that the 4-dimensional
Einstein gravity is recovered on both branes under the approximation
of zero mode truncation.  An approximate formula for the mass of the
lowest massive mode in the scalar-type perturbations is also obtained.
Furthermore, we present a method to evaluate the metric perturbations
taking into account all the KK contributions under a contact
interaction approximation. This approximation is valid if the
source which excites the KK modes is smoothly distributed compared with
the scale corresponding to the mass of the lowest massive mode.

This paper is organized as follows. In Sec.~II we describe the 
background model that we consider, and derive the basic equations 
for the linear perturbations on it. In Sec.~III we discuss the 
mass spectrum of the perturbations. We identify the mode functions 
for the massless degrees of freedom, 
and also derive the formula for the mass and the mode function of 
the lowest massive mode. 
In Sec.~IV the mechanism to recover 
Einstein gravity is explained. This result is expected 
from the notion of the mass spectrum but 
its derivation is not so trivial when we consider the explicit 
construction of metric perturbations. 
We shall find that the contribution from the massive modes 
of the scalar-type perturbations must be also taken into account 
in some sense. For this purpose, we introduce a contact 
interaction approximation. 
In Sec.~V we apply the same technique to the KK modes of 
the tensor-type perturbations to complete the description 
of the linear perturbations of this system. 
Section VI is devoted to summary.

\section{model and basic equations}
We consider linear perturbations of the model proposed by Randall and
Sundrum\cite{RS1} with the stabilization mechanism by Goldberger and
Wise\cite{GW}.  The fields existing on the 5-dimensional spacetime (bulk)
are the usual 5-dimensional gravity with a negative cosmological
constant $\Lambda$, and a 5-dimensional scalar field $\varphi$ 
introduced to stabilize the distance between the two
branes.
We denote the 5-dimensional gravitational constant by $G_5$.  The
unperturbed metric is supposed to take the form,
\begin{equation}
 ds^2 = a^2(y)\eta_{\mu\nu}dx^{\mu}dx^{\nu}+dy^2,
\end{equation}
where $\eta_{\mu\nu}$ is the 4-dimensional Minkowski metric with 
$(-+++)$ signature. 
We use $\gamma_{\mu\nu}=a^2(y)\eta_{\mu\nu}$ to raise  
4-dimensional tensor indices. 
The $y$-direction is bounded by two branes located at $y=y^{(\pm)}$.
On these two branes, ${\bf Z}_2$-symmetry is imposed. 
The Lagrangian for the bulk scalar field is 
\begin{equation}
 {\cal L}=-{1\over 2} g^{ab}\varphi_{,a}\varphi_{,b}
          -V_B(\varphi)-\sum_{\sigma=\pm} V^{(\sigma)}(\varphi)
              \delta(y-y^{(\sigma)}). 
\end{equation}
For most of the present analysis, we do not need to specify the 
explicit form of the potentials 
$V_B(\varphi)$ and $V^{(\pm)}(\varphi)$. 
In the bulk, the background scale factor and the scalar field, 
$(a, \varphi_0)$, must satisfy 
\begin{eqnarray}
 \dot H(y) & = & -{\kappa\over 3}\dot\varphi_0^2(y),\cr
 H^2(y) & = & {\kappa \over 6}\left({1\over 2}
            \dot\varphi_0^2(y) - V_B(\varphi_0(y))-\kappa^{-1}\Lambda\right),\cr
 \ddot\varphi_0(y) &+& 4 H(y) \dot \varphi_0(y) -V_B'(\varphi_0(y))=0,
\end{eqnarray}
where $H(y):=\dot a(y)/a(y)\approx -\sqrt{-\Lambda/6}$ and
$\kappa=8\pi G_5$.  Ordinary matter fields reside on both branes, and
the values of the 4-dimensional vacuum energy on both branes are
adjusted to realize a static background configuration.  This way
  of model construction is different from the standard
  approach\cite{GW} which assumes the model potential and the values
  of the vacuum energy on the branes from the very beginning.  
  In the present approach, we
  obtain a constraint on the model potential by fixing the distance
  between the two branes.  Conversely, in the standard approach, the
  stabilized distance is determined for each given model 
  (although some tuning of one of the model parameters is necessary 
  to realize a static configuration).  We call
the brane at $y=y^{(+)} (y^{(-)})$ the positive (negative) tension
brane, and adopt the convention $y^{(+)}<y^{(-)}$.

We first consider the metric perturbation $\delta(ds^2)=h_{ab}dx^a dx^b$ 
and the scalar field perturbation $\delta\varphi$ in the bulk. 
We are using the convention that Latin indices run through $0,1,2,3,5$, 
while Greek indices run through $0,1,2,3$. 
In the bulk, we can always impose the ``Newton gauge'' condition, 
\begin{eqnarray}
 h_{55} & = & 2\phi^N,\quad h_{5\mu}=0, \cr
 h_{\mu\nu} & = &h_{\mu\nu}^{(TT)}
   -\phi^N\gamma_{\mu\nu},\cr
 \delta\varphi & = & {3\over 2\kappa\dot\varphi_0}
       \left[\partial_y+2H \right]\phi^N, 
\end{eqnarray}
where $h_{\mu\nu}^{(TT)}$ satisfies the transverse-traceless 
condition, and $\phi^N$ is the 5-dimensional ``Newtonian potential''. 

The equation for $h_{\mu\nu}^{(TT)}$ in the bulk is given by 
\begin{equation}
\left[ {1\over a^2}\Box^{(4)}+\hat L^{(TT)}
       \right] h_{\mu\nu}^{(TT)} = 0,
\label{tensor}
\end{equation}
where $\Box^{(4)}$ is the 4-dimensional d'Alembertian operator with 
respect to $\eta_{\mu\nu}$, and we have introduced the operator
\begin{equation}
 \hat L^{(TT)}:={1\over a^2}\partial_y a^4\partial_y {1\over a^2}. 
\end{equation}
For the scalar-type perturbations, we obtain 
\begin{equation}
\left[\Box^{(4)}+\hat L^{(\phi^N)}
       \right]\phi^N = 0,
\label{scalar}
\end{equation}
where $\hat L^{(\phi^N)}$ is the operator defined by 
\begin{equation}
 \hat L^{(\phi^N)}:=a^2\dot\varphi_0^2\partial_y{1\over a^{2}\dot\varphi_0^2}
     \partial_y a^2-{2\kappa\over 3}a^2\dot\varphi_0^2.
\end{equation}

As discussed in Paper I, the $y$ = constant hypersurface does not
correspond to the location of the branes in this gauge. Following it, we
introduce two Gaussian normal coordinate systems near both branes,
respectively.  We denote quantities in these coordinate systems by
associating a bar, such as $\bar{y}$.

Then, the junction condition for the bulk scalar field is given by 
$\pm 2\dot{\bar\varphi}=V'{}^{(\pm)}(\bar\varphi)$ at
$\bar y=\bar y^{(\pm)}\pm$. 
Hence, its perturbation become
\begin{equation}
  \pm 2\delta\dot{\bar\varphi}=
   V''{}^{(\pm)}(\varphi_0){\delta\bar\varphi},\quad
   (\bar y=\bar y^{(\pm)}\pm). 
\label{junctionscalar}
\end{equation}
The junction condition for the metric perturbation 
is
\begin{eqnarray}
 \pm(\partial_{\bar{y}}-2H)\bar h_{\mu\nu}
  &=& -\kappa\left[T_{\mu\nu}-{1\over 3}\gamma_{\mu\nu}T\right]^{(\pm)}\cr
  && \mp{2\kappa\over 3}\gamma_{\mu\nu}\dot\varphi_0
   \delta\bar\varphi,\quad
 (\bar y=\bar y^{(\pm)}\pm).
\label{hjunc}
\end{eqnarray}
Here, we have introduced the energy momentum tensor of the matter
fields confined on the branes, $T_{\mu\nu}^{(\pm)}$, and
$T^{(\pm)}:=\gamma^{\mu\nu}T_{\mu\nu}^{(\pm)}$. $T_{\mu\nu}^{(\pm)}$
satisfies the 4-dimensional conservation law $T_{\mu\nu}{}^{;\nu}=0$.
Notice that there appears a contribution from the perturbation of the
scalar field $\varphi$ which was not present in the models discussed
in Paper~I.

Whereas the junction conditions have been easily derived using
Gaussian normal coordinates, the equations of motion for the
perturbations are simpler in the ``Newton gauge''. So we need to
consider the gauge transformation which relates normal coordinates and
the ``Newton gauge''. This is given by $h_{ab}=\bar h_{ab}
+\xi_{a;b}+\xi_{b;a}$ with
\begin{eqnarray}
 \xi^{5}_{(\pm)} & = &  \int^y \phi^N (y') dy'
        =\int_{y^{(\pm)}}^y \phi^N(y') dy' +\hat\xi^5_{(\pm)}, \cr
 \xi^{\nu}_{(\pm)} & = & -\int^y \gamma^{\mu\nu}(y')
                  dy'\int^{y'} \phi^N_{,\mu}(y'') dy''\\
& =& -\int^y_{y^{(\pm)}} \gamma^{\mu\nu}(y')
                  dy'\int^{y'} \phi^N_{,\mu}(y'') dy''+\hat\xi^\nu_{(\pm)},
\label{gautransf}
\end{eqnarray}
where $\hat\xi^5_{(\pm)}$ and $\hat\xi^\nu_{(\pm)}$ are independent of $y$. 
Then, we have 
\begin{eqnarray}
\delta\bar \varphi(y)
 &=& \delta\varphi(y) -\dot\varphi_0(y) \left[\int_{y^{(\pm)}}^y \phi^N(y') dy' +\hat\xi^5_{(\pm)}\right], \cr
 \bar h_{\mu\nu}(y) & = & h_{\mu\nu}(y)+ 2a^2(y)\int^y {dy'\over a^{2}(y')}
                \int^{y'} \phi^N_{,\mu\nu}(y'') dy''\hspace{-5mm}\cr
   &&\quad\quad\quad 
                 -2H\gamma_{\mu\nu}(y)\int^y \phi^N(y') dy'.
\label{hgauge}
\end{eqnarray}
Here we should stress that the arguments in the l.h.s. are 
not $\bar y$ but $y$.  

Combining Eqs.(\ref{hjunc}) and (\ref{hgauge}), the junction condition
in the ``Newton gauge'' for the $TT$ part is obtained as 
\begin{eqnarray}
 &&\pm(\partial_y-2H) h_{\mu\nu}^{(TT)}
 = -\kappa\Sigma_{\mu\nu}^{(\pm)},
   \quad(y=y^{(\pm)}\pm),
\label{jeqTT}
\end{eqnarray}
where we have defined 
\begin{eqnarray}
\Sigma_{\mu\nu}^{(\pm)} &:=&
 \left(T_{\mu\nu}-{1\over 4}\gamma_{\mu\nu}T\right)^{(\pm)}\cr
  &&\quad \pm {2\over\kappa}\left(\hat\xi^5_{(\pm),\mu\nu}-{1\over 4}\gamma_{\mu\nu}
           \hat\xi_{(\pm)}^5{}_{,\rho}{}^{,\rho}\right). 
\end{eqnarray}
Combining Eq.~(\ref{jeqTT}) with Eq.~(\ref{tensor}), we obtain
the equation for $h_{\mu\nu}^{(TT)}$,
\begin{eqnarray}
&& \left[{\Box^{(4)}\over a^2}+{\hat L^{(TT)}}
         \right] h_{\mu\nu}^{(TT)}
=-2\kappa \sum_{\sigma=\pm} \Sigma_{\mu\nu}^{(\sigma)}
         \delta(y-y^{(\sigma)}). 
\label{TTeq}
\end{eqnarray}
 
The resulting $h_{\mu\nu}^{(TT)}$ is automatically consistent with the
traceless condition, but the transverse condition gives us the
equation which determines $\hat\xi^{5}_{(\pm)}$,
\begin{equation}
 {1\over a^2_{(\pm)}}\Box^{(4)}\hat\xi_{(\pm)}^5=\pm{\kappa\over 6}T^{(\pm)},
\label{eqxi5}
\end{equation}
where we have defined $a_{(\pm)}:=a(y^{(\pm)})$.  The set of
Eqs.(\ref{TTeq}) and (\ref{eqxi5}) is exactly the same that was
obtained in Paper I once we specialize the background bulk geometry to
pure anti de Sitter. 

Using Eq.~(\ref{eqxi5}), it can be easily seen that the trace
part of the metric junction condition is trivially satisfied.

The remaining junction condition is the one for the scalar field,
Eq.~(\ref{junctionscalar}). After some computations, it reduces in the
``Newton gauge'' to
\begin{equation}
 \mp{2 \kappa\over 3}
 (\delta\varphi-\dot\varphi_0\hat\xi_{(\pm)}^5)
 ={\epsilon^{(\pm)}\over a^2\dot\varphi_0}\Box^{(4)} \phi^N,
\quad(y=y^{(\pm)}\pm), 
\label{juncend}
\end{equation}
where we have defined 
\begin{equation}
\epsilon^{(\pm)}
:={2\over V''{}^{(\pm)}\mp 2(\ddot\varphi_0/\dot\varphi_0)}\quad. 
\end{equation}
Combining Eq.~(\ref{juncend}) with Eq.~(\ref{scalar}), we obtain
the equation for $\phi^N$,
\begin{eqnarray}
&& \hat L^{(\phi^N)}\phi^N
 -\sum_{\sigma=\pm} \sigma {4\kappa a^2\dot\varphi_0^2\over 3}
     \hat\xi^5_{(\sigma)}\delta(y-y^{(\sigma)}) \cr
      &&\quad\quad\quad= -\Box^{(4)} \phi^N\left(1+\sum_{\sigma=\pm} 
          2\epsilon^{(\sigma)}\delta(y-y^{(\sigma)})\right). 
\label{eqphiN}
\end{eqnarray}

\section{mass spectrum}
\subsection{Zero modes}
Let us first consider the solution of the source free equations by
setting $T_{\mu\nu}^{(\pm)}=0$.  We will first consider the zero
eigenmodes of $\Box^{(4)}$, which are the most important because they
correspond to the 4-dimensional massless field responsible for the
propagation of a long range force.

If we set $\Box^{(4)}=0$ in Eq.~(\ref{TTeq}), the solution for
$h_{\mu\nu}^{(TT)}$ is
\begin{equation}
 h_{\mu\nu}^{(TT)}
    = \hat h^{(1)}_{\mu\nu}(x^{\rho}) a^2(y)+ \hat h^{(2)}_{\mu\nu}(x^{\rho}) 
    a^2(y)\int^y {dy'\over a^4(y')},
\end{equation}
where $\hat h^{(1)}_{\mu\nu}$ and $\hat h^{(2)}_{\mu\nu}$ are
4-dimensional $TT$ tensors independent of $y$ satisfying
$\Box^{(4)}\hat h_{\mu\nu}^{(i)}=0$ .  Furthermore, the junction condition
(\ref{jeqTT}) gives
\begin{equation}
{\hat h^{(2)}_{\mu\nu}=-2 a^2_{(\pm)}} \hat\xi^5_{(\pm),\mu\nu}. 
\label{xi5c2}
\end{equation}
Hence, there is no extra constraint on $\hat h^{(1)}_{\mu\nu}$,
but $\hat h^{(2)}_{\mu\nu}$ must be written as the second derivative
of a scalar function.  The former mode corresponds to 4-dimensional
gravitational wave perturbations. Although there seem to exist 5
independent degrees of freedom in this mode, three of them are pure
gauge.  On the other hand, the latter mode should be classified as 
scalar-type perturbations.  If we forget about the stabilization
mechanism, this mode corresponds to the mode called radion in
Ref.\cite{Rub}.

Notice that Eq.~(\ref{xi5c2}) gives a relation between $\hat
\xi^5_{(+)}$ and $\hat \xi^5_{(-)}$,
\begin{equation}
  a^2_{(+)}\hat \xi^5_{(+)}=a^2_{(-)}\hat \xi^5_{(-)}.
\label{zeromoderel}
\end{equation} 
In the present case with stabilization, $\hat\xi^5_{(\pm)}$ also
appears in the junction condition for the scalar-type perturbations.
Hence, the scalar field perturbation must also be chosen to be
compatible with this condition.  Here we should note that this
relation holds only when we restrict our considerations to zero modes.
In the general case, which will be considered later, there is no reason for
such a relation to be satisfied.

Let us now consider the solution of the zero eigenvalue
scalar-type perturbation.
One solution is
\begin{eqnarray}
&& \phi^N={H\over a^2} f^{(1)}(x^{\rho}),\cr 
&& \delta\varphi=-{\dot\varphi_0\over 2a^2}f^{(1)}(x^{\rho}).
\label{phiN1}
\end{eqnarray}
This mode satisfies the scalar field junction
condition (\ref{juncend}) with (\ref{zeromoderel}) if
$f^{(1)}=-2a^2_{(\pm)}\hat\xi^5_{(\pm)}$.  However, this solution
can be 
transformed to nothing by a gauge transformation with parameters
$\xi^5=a^{-2}{f^{(1)}/2}$, $\xi^{\mu}=-f^{(1)}_{,\nu}\int^y a^{-2}(y')
\gamma^{\mu\nu}(y'){dy'}$.  

The other solution is
\begin{eqnarray}
&& \phi^N=\left(1 - {2H\over a^2}\int^y a^2(y') dy'\right)
        f^{(2)}(x^{\rho}),\cr
&& \delta\varphi=-f^{(2)}(x^{\rho})
    {\dot\varphi_0 \over a^2}\int^y a^2(y') dy'.
\label{phiN2}
\end{eqnarray}
However, this mode is not compatible with condition
(\ref{zeromoderel}). Hence, if we include the stabilization mechanism,
no physical massless mode is present in the scalar-type perturbation
spectrum, in contrast with the the case discussed in Paper I.

\subsection{Massive modes}
Let us now consider the lowest massive mode.
They are also important because they give the leading order correction
to the zero mode truncation approximation.  Furthermore, if tachyonic
modes are present, such a model must be rejected.

To estimate the lowest mass eigenvalue in the general case, one
needs numerical calculations. In order to obtain analytical
approximations, here we assume that the effect of the 
bulk scalar field back reaction
to the background geometry is small.  Namely,
we assume
\begin{equation}
 {|\dot H|\over H^2}={\kappa\dot\varphi_0^2\over 3H^2}\ll 1. 
\end{equation}
For the metric, we can use the pure anti-de Sitter form
\begin{equation}
 a(y)=e^{-y/\ell}. 
\label{aassume}
\end{equation} 
For simplicity, here we set 
\begin{equation}
 y^{(+)}=0,\quad y^{(-)}=d.
\label{yassume}
\end{equation}

First we consider the tensor-type perturbations.  The expression for the
mode functions $u_i(y)$ for the operator $\hat L^{(TT)}$ satisfying
the junction condition $(\partial_y-2H)u_i(y)=0$ on the positive
tension brane is found in Refs.\cite{RS2,GT}.  Denoting the eigenvalue
corresponding to $u_i(y)$ by $m_i^2$, the mode function is given in
terms of Bessel functions, $ u_i(y)\propto \{J_1(m_i\ell)Y_2(m_i\ell
e^{y/\ell}) - Y_1(m_i\ell) J_2(m_i\ell e^{y/\ell})\}$.  The boundary
condition on the negative tension brane reduces to
\begin{equation}
\left\{J_1\left({m_i\ell}\right)Y_1\left({m_i\ell e^{d/\ell}}\right) - 
Y_1\left({m_i\ell}\right) J_1\left({m_i\ell e^{d/\ell}}\right)\right\}=0,
\end{equation} 
and determines the discrete eigenvalues $m_i$. 
For small $m_i$, this condition becomes 
$J_1(m_i\ell e^{d/\ell})\approx 0$, 
and hence the eigenvalues are given by 
$m_i\approx e^{-d/\ell} j_i \ell^{-1}$, where 
$j_i$ is the $i$-th zero-point 
of $J_1$. 
Hence the physical mass of the lowest massive KK mode 
on the positive tension brane is given by 
$\approx 3.8 e^{-d/\ell} \ell^{-1}$ while that on the negative 
tension brane by $\approx 3.8\ell^{-1}$.

Next, let us consider the scalar-type perturbations.  When we discuss
the scalar-type perturbations, we should not assume an scale factor of
the form (\ref{aassume}) from the beginning, because the non-vanishing
second or higher derivative of $a(y)$ can be important.  Hence we will
derive formula (\ref{massformula1}) for the lowest mass eigenvalue
without assuming (\ref{aassume}).  Only to obtain an estimate of this
expression, we will use the scale factor (\ref{aassume}).

To discuss scalar-type perturbations, it is convenient to 
introduce a new variable defined by 
\begin{equation}
q:=\frac{3a^2}{2\kappa A}\phi^N,
\label{defq}
\end{equation} 
where a prime ${}'$ denotes derivative with respect to the
conformal coordinate $z$ defined by $dz= a(y)^{-1}dy$, and we have
introduced $A=\pm a^{1/2}\varphi'_0$.  The signature in the definition
of $A$ is chosen so that $A$ is continuous on the branes.  The
junction condition for $q$ is obtained from (\ref{juncend}),
\begin{equation}
2[\partial_{z}+(A'/A)]q=\pm 2A\hat\xi^5_{(\pm)}
\quad(y=y^{(\pm)}\pm).
\label{boundaryq}
\end{equation}
The perturbation equation in terms of $q$ becomes 
\begin{eqnarray}
&&\left[\Box^{(4)}+\partial_{z}^2+A\left({1\over A}\right)''
   -{2\kappa\over 3}{\varphi'_0}^2\right]q \cr\quad 
&&\hspace*{2cm} =\sum\pm 2A \hat\xi^5_{(\pm)}\delta(y-y^{(\pm)}).
\label{eqq}
\end{eqnarray}
Integrating the above equation once at the vicinity of the branes, 
we correctly reproduce the junction condition (\ref{boundaryq}).

To obtain the solution of Eq.~(\ref{eqq}), we will construct the
Green function for the differential operator appearing in it.  It is
given by
\begin{equation}
G_q=-\int {d^4k\over(2\pi)^4} e^{ik_{\mu}\Delta x^{\mu}}
    \sum_i{q_i(z)q_i(z')\over m_i^2+k^2},  
\label{greenq}
\end{equation}
where $m_i$ is the mass eigenvalue. The mode functions satisfy 
\begin{equation}
  \left[\partial^2_{z}-A\left({1\over A}\right)''
   -{2\kappa\over 3}{\varphi'_0}^2\right]q_i=
    -m_i^2 q_i,
\label{modeeq}
\end{equation} 
with the boundary condition 
$\left[\partial_{z}+(A'/ A)\right]q_i=0$,
at $y=y^{(\pm)}\pm$.

In the weak back reaction case, the term $\delta
U:={2\kappa}{\varphi'_0}^2/3$ in Eq.~(\ref{modeeq}) is assumed to be
small. If we perturbatively expand the mode function and the mass
eigenvalue with respect to powers of $\delta U$ like
$q_i=q_i^{(0)}+q_i^{(1)}+\cdots$ and $m_i=m_i^{(0)}+m_i^{(1)}+\cdots$,
the lowest eigenmode for the unperturbed system is trivially given by
\begin{equation}
 q_0^{(0)}={{\cal N}\over A}, 
\end{equation}
with $(m_0^{(0)})^2=0$, where ${\cal N}$ is a normalization 
constant. 

It is straightforward to obtain the next order correction. We find
\begin{eqnarray}
q_0^{(1)} & = & {{\cal N}\over A}\int^{z} A^2 dz'
           \int^{z'}{dz''\over A^2}
           \left({2\kappa\over 3}\varphi'_0{}^2 -(m_0^{(1)})^2\right).  
\end{eqnarray}
In terms of $\delta\varphi$, the above first order correction is given
by
\begin{equation}
 \delta\varphi_0^{(1)}=
    {\cal N}\dot\varphi_0(y)
        \int_{y^{(+)}}^{y}{dy'}
           \left({2\kappa \over 3a^2} 
         -{(m_0^{(1)})^2\over a^4(y')\dot\varphi_0^2(y')} \right). 
\end{equation}
The junction condition on the positive tension brane is already
imposed due to the choice of the integration constant.  Here, for
simplicity, we have taken the $\epsilon^{(\pm)}\to 0$ limit.

This expression is seen to be regular even when $\dot\varphi_0$
vanishes at some points. Denoting the point of vanishing
$\dot\varphi_0$ by $y_0$, we can show that $a^4\dot\varphi_0^2$ can be
expanded around this point as
\begin{eqnarray}
 a^4\dot\varphi_0^2 = \left[a^4 (V'_B)^2\right]_{y=y_0}
           \Bigl[&&(y-y_0)^2\cr
 &&+ \alpha^2
(y-y_0)^4+\cdots\Bigr],  
\label{expand}
\end{eqnarray}
where $\alpha^2:={1\over 3}(4H^2+V''_B)|_{y=y_0}$.  
From this, it is easy
to see that the expression for $\delta\varphi^{(1)}_0$ is regular
at $y=y_0$.

The junction condition (\ref{boundaryq}) on the negative tension
brane gives the formula for the lowest mass eigenvalue,
\begin{equation}
 m_0^2 
 \approx {2\kappa\over 3}\left.\left[\int_{y^{(+)}}^{y^{(-)}} {dy\over a^2}\right]
        \right/ \left[\int_{y^{(+)}}^{y^{(-)}} {dy\over a^4\dot\varphi_0^2}\right]. 
\label{massformula1}
\end{equation}
This expression is positive definite if there is no point at which
$\dot\varphi_0$ vanishes.  However, when $\dot\varphi_0$ vanishes, the
integral in the denominator changes its signature at that point.
Hence, $m_0^2$ can be negative.  As seen from Eq.~(\ref{expand}),
 the integrand does not have residue at this point.  Hence the
 integral is independent of the way of modification of the
 integration path.  It is expected that the dominant contribution to
this integral comes from the region near the point $y_0$, where
$\dot\varphi$ vanishes.  By substituting equation (\ref{expand}), we
can approximately evaluate the denominator in Eq.~(\ref{massformula1})
as
\begin{eqnarray*}
&&\approx
   \int dy{\left[a^4 (V'_B)^2\right]_{y=y_0}^{-1}
    \over (y-y_0)^2+\alpha^2 (y-y_0)^4}
=-\left[\pi\alpha \over a^4 (V'_B)^2\right]_{y=y_0}, 
\end{eqnarray*}
which becomes negative.  Although we cannot give a proof here,
this seems to be the case in general.  Here we simply conjecture it.
If tachyonic modes appear, they mean the breakdown of the model.
Hence, we do not consider the case in which $\dot\varphi_0$ vanishes
at some point, and hereafter we assume that $\dot\varphi_0$ has a
definite sign.

Without explicitly specifying any model, we can make a crude
estimate of the lowest mass eigenvalue. To evaluate the numerator in
the r.h.s of Eq.~(\ref{massformula1}), we can use the form
(\ref{aassume}) for the scale factor $a$.  To evaluate the
denominator, we use the following fact.  Besides some exceptional
cases, $a^4\dot\varphi_0^2$ will have a minimum (cases with no
minima will be discussed later). There
$\ddot\varphi_0+2H\dot\varphi_0=V'_B-2H\dot\varphi_0=0$, and
accordingly the integrand has a rather sharp peak.
We denote this point by $y_c$.  At this point, the second derivative
of $a^4\dot\varphi_0^2$ is evaluated as $(8H^2+2V''_B-4\dot
H)a^4\varphi_0^2$.  Hence, we can approximate the integral like $\int
(1/ a^4\dot\varphi_0^2) dy \approx [{4H^2/ a^4 (V'_B)^2}]_{y=y_c} \int
\exp[ -(4H^2+V''_B)_{y=y_c}(y-y_c)^2]dy$.  Under these approximations,
the formula for the mass of the lowest eigenmode is obtained as
\begin{eqnarray}
 m_0^2
  \approx
  {\kappa e^{2d/\ell}\ell^2\over 6\sqrt{\pi}} 
 \left[e^{-4y/\ell} (V'_B)^2 \sqrt{1+{V''_B \ell^2\over 4}}\right]_{y=y_c}.
\label{massf}
\end{eqnarray}

To proceed further, we consider the specific model for the bulk
potential $V_B(\varphi)$ discussed in Ref.\cite{GW},
\begin{equation}
  V_B(\varphi)={M^2\varphi^2\over 2}.
\end{equation} 
For this model, 
the background solution for the bulk scalar field 
in the weak back reaction case is already given by 
\begin{equation}
\varphi_0=B_1 e^{\nu_1 y}+B_2 e^{\nu_2 y},
\end{equation}
where $\nu_1=2\ell^{-1}+\sqrt{4\ell^{-2}+M^2}$, 
$\nu_2=2\ell^{-1}-\sqrt{4\ell^{-2}+M^2}$
and 
\begin{eqnarray}
 && B_1\approx e^{-\nu_1 d}
     \left(\varphi_{(-)}-\varphi_{(+)}e^{\nu_2 d}\right),\cr 
 && B_2\approx \varphi_{(+)}-\varphi_{(-)}e^{-\nu_1 d}.
\end{eqnarray} 
$\varphi_{(+)}$ and $\varphi_{(-)}$ are the values of $\varphi_0$ on
the positive and the negative tension branes, respectively.  

Let us briefly elaborate on the stabilization distance. For simplicity, we
will assume that the ratio $\varphi_{(+)}/\varphi_{(-)}$ is not
extremely large, and also that all the input scales are similar, i.e.
$\kappa\approx\ell^3$. The condition for weak back reaction is
\begin{equation}
\dot\varphi_0^2\ll {1\over \kappa\ell^2}. 
\end{equation}
Since $\dot\varphi_0^2$ does not have its maximum in the bulk,
it is sufficient to consider the conditions
for weak back reaction on the boundaries. Then, we can evaluate the
values of $\dot\varphi_0$ on boundaries,
\begin{eqnarray}
&&\dot \varphi_0 (y^{(+)})\approx \nu_2\varphi_{(+)},\cr
&&\dot \varphi_0 (y^{(-)})\approx \nu_1\varphi_{(-)}
               -4\ell^{-1}\sqrt{1+(M^2\ell^2/4)}
              e^{\nu_2 d}\varphi_{(+)}. 
\end{eqnarray}
The conditions for weak back reaction on 
both branes become
\begin{eqnarray}
 &&\left({\nu_2\varphi_{(+)}\over 4\ell^{-5/2}}\right)^2
    \ll {\ell^3\over\kappa},\cr
 && \left({\nu_1\varphi_{(-)}\over 4\ell^{-5/2}}\right)^2
  \left(1-{4e^{\nu_2 d}\over \nu_1\ell}
             {\varphi_{(+)}\over \varphi_{(-)}}
                        \sqrt{1+{M^2\ell^2\over 4}}\right)
  \ll {\ell^3\over\kappa}.
\end{eqnarray}
The condition on the positive tension brane is satisfied by choosing
({\it case A}) $\varphi_{(\pm)}\ll\ell^{-3/2}$ with $M^2\alt\ell^{-2}$
or ({\it case B})
$M^2\ll\ell^{-2}$ with $\varphi_{(\pm)}\agt \ell^{-3/2}$. 

In case A, the condition on the negative
tension brane is automatically satisfied. In this case, the
stabilization distance is sensitive to the change of the values of
the vacuum energy on the branes. Hence, without specifying the complete
model, we cannot estimate the stabilization distance.
If we consider the case in which $M^2\ell^2$ is not small, 
we find that it is natural to assume that the signature 
of $\varphi_{(+)}$ is different from that of $\varphi_{(-)}$. 
If $\varphi_{(+)}$ and $\varphi_{(-)}$ have the same signature, 
the ratio between them must be chosen to be extremely large 
to find a node-less solution of 
$\dot\varphi_0$ with sufficiently large brane separation $d$.
In the case that $\varphi_{(+)}$ and $\varphi_{(-)}$ have 
different signature, the solution of $\dot\varphi_0$ becomes 
node-less for any choice of parameters. 

Case B is the case discussed in Ref.\cite{GW}. 
In this case, to satisfy the
condition for weak back reaction on the negative tension brane, 
\begin{equation}
 {d \over \ell}\approx {1\over -\nu_2\ell}\ln\left(
       {4\ell^{-1}\varphi_{(+)}
     \sqrt{1+{M^2\ell^2\over4}} \over \nu_1\varphi_{(-)}}\right) 
\end{equation}
is required. 
Hence, $\varphi_{(+)}$ and $\varphi_{(-)}$ 
must have the same signature.  To realize a sufficiently 
large value of $d/\ell$, it is necessary that 
$|\varphi_{(+)}|$ is slightly larger than $|\varphi_{(-)}|$. 
If we take the small $M^2$ limit, the above expression 
for the stabilization distance coincides with 
that obtained in Ref.~\cite{GW}.

Let us now return to the computation of $m_0^2$. 
First we consider the case in which $a^4\dot\varphi_0^2$ 
has a minimum and the estimate (\ref{massf}) is valid. 
As we shall see below, case B is exceptional in this sense. 
From the condition
that $V'_B-2H\dot\varphi_0=0$ at $y=y_c$, we have
$e^{(\nu_2-\nu_1)y_c}={\nu_1 B_1/\nu_2 B_2}$.  Substituting this
estimate of $y_c$ into (\ref{massf}), we obtain
\begin{equation}
 m_0^2
  \approx{8\kappa M^2 e^{2d/\ell}\sqrt{1+(M^2\ell^2/4)}\over 3\sqrt{\pi}} 
     (-B_1 B_2).
\label{massf2}
\end{equation}
 
To solve the hierarchy problem on the negative tension brane, we need
to set $e^{d/\ell}\sim 10^{19}$(GeV)$/10^{3}$(GeV)$=10^{16}$.  Hence,
$d/\ell\approx 37$.  Therefore, it is rather natural to suppose that
$M^2\ell d$ is larger than unity.  If $|\varphi_{(+)}|$ is not
much larger than $|\varphi_{(-)}|$, as assumed, we can approximate
$-B_1 B_2\approx |\varphi_{(+)}\varphi_{(-)}| e^{-\nu_1 d}$.  In this
case, the mass eigenvalue $m_0^2$ becomes $O({\kappa |\varphi_{(+)}
  \varphi_{(-)}| M^2} e^{-2\ell^{-1}d\sqrt{1+(M^2\ell^2/4)}})$.  Hence
the physical squared mass on the negative tension brane becomes
$O(\kappa |\varphi_{(+)} \varphi_{(-)}| M^2 e^{-2\ell^{-1}d
  (\sqrt{1+(M^2\ell^2/4)}-1)})$ while that on the positive tension
brane is the same as $m_0^2$.
As for $|\varphi_{(\pm)}|$, they must be smaller than
$\ell^{-3/2}$ for the approximation of weak back reaction to be valid.
Also, there is a factor 
$ M^2 e^{-2\ell^{-1}d (\sqrt{1+(M^2\ell^2/4)}-1)}$, 
which takes its maximum value $(0.2/\ell)^2$ at $M\approx 0.33\ell^{-1}$.   
Hence, the mass scale on the negative
tension brane tends to be smaller than the typical background energy 
scale $\ell^{-1}$.  

Finally, we consider the special case in which $a^4\dot\varphi_0^2$ in
Eq.~(\ref{massformula1}) has no minima.  This happens when one of the
terms in $\dot\varphi_0=\nu_1B_1 e^{\nu_1 y}+\nu_2B_2 e^{\nu_2 y}$
can be totally neglected.  This condition becomes
\begin{equation}
 |\varphi_{(-)}-\varphi_{(+)}e^{\nu_2 d}|\ll
      \left\vert{\nu_2\over \nu_1}\right\vert |\varphi_{(-)}|, 
\end{equation}
or 
\begin{equation}
 |\varphi_{(+)}-\varphi_{(-)}e^{-\nu_1 d}|\ll
      \left\vert{\nu_1\over \nu_2}\right\vert |\varphi_{(+)}|. 
\end{equation}
In the former case we can set $B_1\approx 0$, and  
in the latter case we can set $B_2\approx 0$.  
Then, for the former case, 
formula (\ref{massformula1}) gives 
\begin{equation}
 m_0^2={4\kappa\over 3}\nu_2^2\varphi_{(-)}^2 e^{-2d/\ell}
\left[{\sqrt{1+(M^2\ell^2/4)}\over 
  1-e^{-4\sqrt{1+(M^2\ell^2/4)} d/\ell}}\right].  
\end{equation}
It is easy to see that case B 
corresponds to this case.  The model
discussed in Sec. 4.1 of Ref.\cite{DeW} with negative small value of
$b$ also corresponds to this case (for the definition of $b$, see
Ref.\cite{DeW}).  The factor in the square brackets is almost unity.
Substituting the approximation $\nu_2\approx -M^2\ell/4$, we recover
the result obtained in \cite{GW}. Note that their $M^3$ and $k$ are
our $1/(4\kappa)$ and $\ell^{-1}$, respectively.

For the latter case, we have 
\begin{equation}
 m_0^2={4\kappa\over 3}\nu_1^2\varphi_{(-)}^2 e^{-2d/\ell}
  \left[{\sqrt{1+(M^2\ell^2/4)}\over 
   e^{4\sqrt{1+(M^2\ell^2/4)} d/\ell}-1}\right]. 
\end{equation}
The model discussed in Sec. 4.1 of Ref.\cite{DeW} with positive 
small value of $b$ corresponds to this case. 
To keep the mass 
sufficiently large, we have to choose $|\varphi_{(-)}|$ extremely large, 
which is not compatible with the weak back reaction condition.   


\section{Recovery of Einstein gravity}

As we have shown in the preceding section, in two-brane models
with stabilization mechanism, a physical massless mode is absent in
the scalar-type perturbation spectrum, which is different from
what happens in the case without stabilization mechanism discussed in
Paper I.  This fact indicates that the resulting 4-dimensional
effective gravity can resemble Einstein gravity at linear order. Let
us show how it can be recovered. 

To compute the induced metric on the branes, it is convenient to
transform back to Gaussian coordinates. Thus, applying ``minus'' the gauge
transformation (\ref{gautransf}), the induced metric on each brane is given by
\begin{eqnarray}
\bar h_{\mu\nu}^{(\pm)} & =& h_{\mu\nu}^{(0)}(y^{(\pm)})
           + h_{\mu\nu}^{(KK)}(y^{(\pm)})\cr
          && \hspace{-5mm}-\gamma_{\mu\nu}\left(\phi^N(y^{(\pm)})
                 +2H\hat\xi^5_{(\pm)}\right)
          -(\hat\xi^{(\pm)}_{\mu,\nu}+\hat\xi^{(\pm)}_{\nu,\mu}).
\label{hbar}
\end{eqnarray} 
Here we have decomposed $h_{\mu\nu}^{(TT)}$ into two parts, the zero
mode contribution $h_{\mu\nu}^{(0)}$ and the KK contribution
$h_{\mu\nu}^{(KK)}$.  The last term represents the residual
4-dimensional gauge transformation. As we have already noted,
Eq.~(\ref{TTeq}) and Eq.~(\ref{eqxi5}) determining
$h_{\mu\nu}^{(TT)}(y^{(\pm)})$ and $\hat\xi^5_{(\pm)}$ are essentially 
the same as the ones obtained in Paper I for the case without stabilization
mechanism. There, it was found that the induced gravity approximated
by the zero mode truncation becomes of the Brans-Dicke type\cite{GT}.
Here we only quote the result for $h_{\mu\nu}^{(0)}(y^{(\pm)})$ 
up to 4-dimensional gauge
transformation,
\begin{eqnarray}
 h_{\mu\nu}^{(0)}(y^{(\pm)}) 
      & = & -\left[{\Box^{(4)}\over a^2}\right]^{-1} \sum_{\sigma=\pm}
      16\pi G^{(\sigma)}\left[T_{\mu\nu}
        -\gamma_{\mu\nu}{T\over 3}\right]^{(\sigma)},\nonumber\\
\label{quote}
\end{eqnarray}
where $8\pi G^{(\pm)}:=\kappa N a^2_{(\pm)}$ and
\begin{equation}
 N:=\left[2\int_{y^{(+)}}^{y^{(-)}} a^2 dy\right]^{-1}.
\end{equation}
Hence, the only possible source of an additional contribution
is that from $h_{\mu\nu}^{(KK)}$ or $\phi^N(y^{(\pm)})$. 
Although we have just shown 
that there are no massless degrees of freedom in the 
scalar-type perturbations, we will find that 
the contribution from $\phi^N(y^{(\pm)})$ gives
the correct long range force required to recover 
Einstein gravity. 

First we consider the weak back reaction case 
discussed in Sec.~III-B. 
From the normalization condition of $q_0^{(0)}$ we obtain \
${\cal N}^2=\left[2\int_0^d dz \left({1/ A}\right)^2\right]^{-1}$.
Comparing it with the formula for the mass (\ref{massformula1}), 
we find a simple relation
\begin{equation}
 m_0^2\approx {4\kappa {\cal N}^2\over 3}\int_0^d {dy\over a^2}
 \approx {2\kappa\ell {\cal N}^2\over 3}(e^{2d/\ell}-1). 
\end{equation}

Then, taking into account the contribution from the modes with the
lowest mass eigenvalue alone, the Green function for $q$, Eq.~(\ref{greenq}),
is approximated by
\begin{eqnarray}
G_q &\approx & 
 -\int {d^4k\over(2\pi)^4} 
    {3e^{ik_{\mu}\Delta x^{\mu}} \over 2\kappa\ell(e^{2d/\ell}-1)} {m_0^2 A^{-1}(y)
     A^{-1}(y')\over m_0^2+k^2}\cr
&= & 
 -    {3A^{-1}(y)A^{-1}(y')\over 2\kappa\ell(e^{2d/\ell}-1)}\cr 
&& \quad      \times \Biggl[\delta^4(\Delta x^{\mu})
         -\int{d^4k\over(2\pi)^4} 
         {k^2 e^{ik_{\mu}\Delta x^{\mu}} \over m_0^2+k^2}\Biggr].
\label{Gqapprox}
\end{eqnarray}
Using this Green function, and with the aid of (\ref{defq}) and (\ref{eqq}), 
we find that $\phi^N$ is given by 
\begin{eqnarray}
 \phi^N&\approx& {2\ell^{-1}\over
              a^2(e^{2d/\ell}-1)}
    \Bigl[(-\hat\xi^5_{(+)}+\hat\xi^5_{(-)})\cr
&&              -{\kappa\over 6}[m_0^2-\Box^{(4)}]^{-1}
                  (a^2_{(+)}T^{(+)}+a^2_{(-)}T^{(-)})\Bigr] . 
\label{phiN}
\end{eqnarray}
Only the first term inside the square brackets gives a long range
contribution to the induced metric.  The propagation of this force is
essentially due to $\hat\xi^5_{(\pm)}$.  The source for the second
term exactly becomes the trace of the energy momentum tensor of the
ordinary matter field.  Hence, the force due to this term becomes
short-ranged with typical length scale $\sim a m_0^{-1}$.  For the
first term, using Eq.~(\ref{eqxi5}) for $\hat\xi^5_{(\pm)}$,
we finally obtain the contribution to the long range component of
$\bar h_{\mu\nu}^{(\pm)}$ coming from $\phi^N$ as 
\begin{eqnarray}
 -\gamma_{\mu\nu}&\phi^N &(y^{(\pm)})
  \approx {8\pi G^{(\mp)}\over 3} 
\cr && \times \left[{\Box^{(4)}\over a^2_{(\pm)}}\right]^{-1}
\left(a^2_{(+)} T^{(+)}+a^2_{(-)} T^{(-)}\right)
       \eta_{\mu\nu},  
\label{correction}
\end{eqnarray}
where we have used the fact that $8\pi G^{(\pm)}=\kappa \ell^{-1}
a_{(\pm)}^2/(1-e^{-2d/\ell})$ in the weak back reaction case.  
Substituting this into Eq.~(\ref{hbar}), and using Eq.~(\ref{quote})
and Eq.~(\ref{eqxi5}), the zero mode truncation reproduces the formula
for the linearized Einstein gravity,
\begin{equation}
\frac{\Box^{(4)}}{a_{(\pm)}^2}\bar h_{\mu\nu}^{(\pm)}= \sum_{\sigma=\pm}
      16\pi G^{(\sigma)}\left[T_{\mu\nu}
        -\gamma_{\mu\nu}{T\over 2}\right]^{(\sigma)}.
\end{equation}

In the above derivation of Eq.~(\ref{correction}), 
we have used several
approximations.  This derivation has the merit of having a result
with a rather easy intuitive interpretation.  The gravitational field
is propagated through the massless field $\hat\xi^{(5)}_{(\pm)}$. At a
point far from the source $T_{\mu\nu}$, $\hat\xi^{(5)}_{(\pm)}$
generates a cloud of metric perturbations through the interaction with
the massive KK modes.  However, the above approximate derivation 
is not completely satisfactory because some aspects of general
relativity are already tested with very high precision.  Hence, we
present an alternative and more complete treatment below.

First we consider to apply the same trick used in the second line of
Eq.(\ref{Gqapprox}) to the complete Green function containing all
massive modes.  Then the long range part of the Green function reduces
to $ G_q\approx -\delta^4(\Delta x^{\mu}) \sum_i{q_i(z)q_i(z')/
  m_i^2}$.  This replacement is exactly valid when we focus on the
long range force.  Now one can notice that to use this Green function
neglecting the terms corresponding to the short range force is
equivalent to solve the equation for $\phi^N$ (\ref{eqphiN}) by
setting $\Box^{(4)}=0$ from the beginning.

{}For the case with $\Box^{(4)}=0$, we have already obtained 
the general
solutions to Eq.~(\ref{eqphiN}), i.e., Eqs.(\ref{phiN1}) and
(\ref{phiN2}).  For convenience, here we quote the previous results in
a slightly different notation, namely
\begin{equation}
 \phi^N=\sum_{\sigma=\pm} u_{\sigma} f^{(\sigma)}(x^{\rho}), 
\end{equation}
with
\begin{equation}
  u_{\pm}:= 1 - {2H\over a^2}\int_{y^{(\mp)}}^y a^2(y') dy'.
\end{equation}
Then the junction condition (\ref{juncend}) with $\Box^{(4)}=0$ 
determines $f^{(\pm)}$ 
as $f^{(\pm)} = \mp 2Na^2_{(\pm)}\hat\xi^5_{(\pm)}$. 
Substituting these back into the expression for $\phi^N$, 
we obtain 
\begin{equation}
 \phi^N(y^{(\pm)})=-2H\hat\xi^5_{(\pm)}
   -2N\sum_{\sigma=\pm}\left[\sigma a^2_{(\sigma)}\hat\xi^5_{(\sigma)}
     \right]. 
\label{phiNlow}
\end{equation}
Adding this contribution to the contributions coming from the
zero mode $TT$ part and $\hat\xi_{(\pm)}^5$ in Eq.~(\ref{hbar}) for the
induced metric on the branes, Einstein gravity at the linear order is
recovered.

It will be illustrative to give the next order correction. 
The source term for the next order correction is 
$-\Box^{(4)}\phi^N_0\left(1+\sum_{\sigma=\pm} 
        2\epsilon^{(\sigma)}\delta(y-y^{(\sigma)})\right)$, 
where we have denoted 
the solution of the lowest order 
approximation (\ref{phiNlow}) by $\phi^N_0$.
By using Eq.~(\ref{eqxi5}), 
$\Box^{(4)}\phi^N_0$ is evaluated as 
\begin{equation}
\Box^{(4)}\phi^N_0 ={\kappa N\over 3}\sum_{\sigma=\pm}u_{\sigma}(y)
           a^{4}_{(\sigma)} T^{(\sigma)}. 
\end{equation}

Then, by using the standard Green function method, 
we obtain 
\begin{eqnarray}
 \phi^N_1(y) & =& 
     u_+\left[\int_{y^{(+)}}^y {u_-\over\dot\varphi^2_0}
     \left({3N\over \kappa}\Box^{(4)}\phi^N_0\right) dy' 
     +K_+\right]\cr 
     &&- u_-\left[\int_{y^{(-)}}^y {u_+\over\dot\varphi^2_0}
     \left({3N\over \kappa}\Box^{(4)}\phi^N_0\right) dy' 
     -K_-\right],  
\end{eqnarray}
where $K_+$ and $K_-$ are integration constants determined 
from the junction condition as
\begin{eqnarray}
 K_{\pm} &= &\left[{N^2\over\dot\varphi^2}\right]^{(\pm)}\epsilon^{(\pm)}
\cr &&
 \times \Biggl[\left\{\left(1\pm{H\over a^2 N}\right)a^4 T\right\}^{(\pm)} 
           +a^4_{(\mp)}T^{(\mp)}\Biggr]. 
\end{eqnarray}
For simplicity, we take again the $\epsilon^{(\pm)}\to 0$ limit.  
Then, we have $K_{\pm}=0$, and we obtain 
\begin{equation}
   \phi^{N(\pm)}_1 
     = N^2 \sum_{\sigma=\pm}
        I_{\pm,\sigma}a^4_{(\sigma)} T^{(\sigma)},
\label{deltaphiN} 
\end{equation}
where we have defined 
\begin{equation}
 I_{i,j}:=\int_{y^{(+)}}^{y^{(-)}} {u_i(y) u_j(y)\over \dot\varphi_0^2} dy. 
\label{Idef}
\end{equation}

In the weak back reaction case, 
we can approximately evaluate this integral by using the 
following facts. 
First, we note that the combination $a^2 u_{\pm}$ can be shown to be a 
slowly changing function for $y\agt y_c$ in the weak back reaction case. 
As before, $y_c$ is the peak location of 
$(a^4\dot\varphi^2_0)^{-1}$. 
To show this, we use the fact that $\dot\varphi_0$ is 
approximated by $\nu_1 B_1 e^{\nu_1 y}$ for $y\agt y_c$. 
This constancy of $a^2 u_{\pm}$ indicates 
that the integrand of Eq.~(\ref{Idef}) 
is approximately proportional to $(a^4\dot\varphi^2_0)^{-1}$. 
Then by using the background geometry defined by 
Eqs.~(\ref{aassume}) and (\ref{yassume}), 
$I_{+,+}$ is evaluated like 
\begin{eqnarray*}
  I_{+,+} &\approx &
    \left[a^2 u_+\right]_{y=d}^2
    \int_{0}^{d} {dy\over a^4 \dot\varphi_0^2}
    \approx {\kappa\ell\over 4m_0^2} e^{2d/\ell}, 
\end{eqnarray*}
where we have used (\ref{massformula1}). 
The other components are also evaluated in a similar way to 
obtain the relations 
$I_{+,+}\approx e^{-2d/\ell} I_{+,-}\approx e^{-4d/\ell} I_{-,-}$. 
The correction obtained by substituting these estimates for $I_{i,j}$ into 
(\ref{deltaphiN}) is consistent with the contribution from 
the second term in the square brackets in Eq.~(\ref{phiN}).  
If we set $\Box^{(4)}=0$ there, we recover the same result.  

It is easy to check that 
the treatment of taking $\Box^{(4)}$ to be small is consistent 
if the distribution of the energy momentum tensor is sufficiently smooth. 
In the treatment presented here, we have taken into 
account the contribution from all the massive modes simultaneously. 
The approximation is completely valid as long as we consider 
smooth matter distribution as compared with the mass 
scale of the lowest massive mode.  
However, the lowest mass on the positive tension brane 
becomes very small if we consider the case in which we are living on 
the negative tension brane. 
In this sense, the treatment presented here is rather restrictive 
when we discuss perturbations caused by the 
matter fields on the positive tension brane.  

\section{TT part Revisited}
Applying the same technique that we have used in 
the preceding section for the scalar-type 
perturbations, we can also deal with the KK mode contribution 
for the $TT$ part $h_{\mu\nu}^{(TT)}$. 
By using the Green function method, we can 
easily calculate the zero mode contribution like 
\begin{equation}
 h_{\mu\nu}^{(0)}=-2N\kappa a^2(y)(\Box^{(4)})^{-1}
         \sum_{\sigma=\pm} a^2_{(\sigma)}\Sigma_{\mu\nu}^{(\sigma)}. 
\end{equation}
After using Eq.~(\ref{eqxi5}), we can recover Eq.~(\ref{quote}).
Substituting $h_{\mu\nu}^{(TT)}=h_{\mu\nu}^{(0)}+h_{\mu\nu}^{(KK)}$
into (\ref{TTeq}), we obtain the equation for the KK contribution,
\begin{eqnarray}
&& \left[a^{-2}\Box^{(4)}+{\hat L^{(TT)}}
        \right] h_{\mu\nu}^{(KK)}
\cr&&\quad\quad\quad
=2\kappa \sum_{\sigma=\pm}
           \Sigma_{\mu\nu}^{(\sigma)}\left[
          a^2_{(\sigma)}N-\delta(y-y^{(\sigma)})\right].
\end{eqnarray}
As before, neglecting the $\Box^{(4)}$-term for the massive KK contribution,
we can solve this equation like
\begin{eqnarray}
  h_{\mu\nu}^{(KK)}
   &=&2N\kappa \sum_{\sigma=\pm}a^2_{(\sigma)}\Sigma_{\mu\nu}^{(\sigma)}
   a^2(y)\cr
   &&\times\left(\int^y_{y^{(-\sigma)}}{dy'\over a^4(y')} 
            \int^{y'}_{y^{(-\sigma)}} dy'' a^2(y'')-C_{(\sigma)}\right),
\label{hmunuKK}
\end{eqnarray}
where $C_{(+)}$ and $C_{(-)}$ are constants. We should recall that
we have already subtracted the zero mode contribution.  Hence
$h^{(KK)}_{\mu\nu}$ must be orthogonal to the zero mode. This 
requires
\begin{equation}
\int_{y^{(+)}}^{y^{(-)}} {dy\over a^2} u_0(y) h_{\mu\nu}^{(KK)}(y) 
 \propto\int_{y^{(+)}}^{y^{(-)}} dy\, h_{\mu\nu}^{(KK)}(y)=0. 
\label{zerosub}
\end{equation}
The constants $C_{(+)}$ and $C_{(-)}$ in the solution (\ref{hmunuKK})
are determined by imposing this condition.

For simplicity we again adopt (\ref{aassume}) with (\ref{yassume}) as
the background geometry.  Then, from condition (\ref{zerosub}),
$C_{(\pm)}$ are explicitly calculated, and the resulting KK
contribution becomes
\begin{eqnarray}
 &&h_{\mu\nu}^{(KK)}
   ={\kappa\ell \over 4(1-e^{-2d/\ell})} 
   \sum_{\sigma=\pm} \Sigma_{\mu\nu}^{(\sigma)}\cr 
 &&\quad \times\Biggl[e^{(2y-2d)/\ell}
          -2a^2_{(\sigma)}+e^{-2y/\ell}
         \left({4a^2_{(\sigma)}d\ell^{-1}\over 1-e^{-2d/\ell}}-1\right)\Biggr].
\end{eqnarray}
Hence, on the respective branes, the contributions from the KK modes 
become
\begin{eqnarray}
 &&  h_{\mu\nu(+)}^{(KK)}
   \approx -{\kappa\ell\over 4} \left(
     (3-4d/\ell)\Sigma_{\mu\nu}^{(+)}+ \Sigma_{\mu\nu}^{(-)}
     \right),\cr
 &&  h_{\mu\nu(-)}^{(KK)}
   \approx {\kappa\ell\over 4} 
        \left(-\Sigma_{\mu\nu}^{(+)}+\Sigma_{\mu\nu}^{(-)}\right),
\label{correctionKK}
\end{eqnarray}
where we have assumed $d/\ell \gg 1$. Here we should recall the 
remaining degrees of freedom for the 4-dimensional gauge 
transformation. 
Using these degrees of freedom, 
$\Sigma_{\mu\nu}$ can be replaced with $T_{\mu\nu}-\gamma_{\mu\nu}
T/3$ with the aid of Eq.~(\ref{eqxi5}).  
Hence, the KK modes, of course, do not give any 
long range force contribution. If one takes the $d/\ell\to\infty$ limit,
$h_{\mu\nu(+)}^{(KK)}$ seems to diverge. But this is just due to the
breakdown of the approximation.  In this limit, the mass difference of 
the KK modes becomes zero.

\section{summary}
In this paper we have developed a systematic procedure to evaluate
the perturbations in the 5-dimensional brane world model proposed by
Randall and Sundrum\cite{RS1} supplemented with the moduli stabilization
mechanism by Goldberger and Wise\cite{GW}.

We have first investigated in detail the mass spectrum of this model.
In the case without stabilization mechanism, there was a scalar-type
massless mode, which was called radion in Ref.\cite{Rub}.  
We have clarified how this
massless mode disappears once we 
switch on the stabilization mechanism.  

We have also estimated the mass eigenvalue and the mode
function corresponding to the lowest mass eigenmodes for both 
tensor-type and scalar-type perturbations, assuming that the back
reaction to the background geometry due to the bulk scalar field
introduced for the moduli stabilization is not large. 
The physical mass of
the lowest tensor-type mode becomes $\approx \ell^{-1}$ on the
negative tension brane and $\approx e^{-d/\ell} \ell^{-1}$ on the
positive tension brane.  Here $\ell$ is the curvature scale of the
background geometry, and $d$ is the proper distance between the two
branes. In the original model, $\ell^{-1}$ is supposed to be TeV
scale.  

{}For the physical mass of the lowest scalar-type mode, 
we have obtained rather general formulas, 
(\ref{massformula1}) and (\ref{massf}). 
To proceed further in estimating the mass, we have  
specified a model for the bulk potential of the scalar field. 
We have considered a simple quadratic potential whose mass is given 
by $M$, and we have assumed that the 5-dimensional gravitational 
constant is also $\approx \ell^3$.  
We pointed out that in this model there are two regimes 
in which the weak back reaction condition holds. 
The first case is the one in which  
the vacuum expectation values of the scalar field on the branes, 
$\varphi_{(\pm)}$, are sufficiently small compared with the background 
energy scale, i.e., $\varphi_{(\pm)}\ll \ell^{-{3/2}}$. 
In this case, we 
found that the mass of the lowest scalar-type mode becomes  
$\approx \sqrt{|\varphi_{(-)}\varphi_{(+)}|\ell^3} 
  M e^{-(\sqrt{1+(M^2\ell^2/4)}-1) d/\ell}$. 
(For a more precise formula, see Eq.~(\ref{massf2}).)  
If we take into account the fact that we need to
set $d/\ell\approx 37$ to solve the hierarchy problem 
on the negative tension brane, this factor is at
most $\approx 0.2\ell^{-1} \sqrt{|\varphi_{(-)}\varphi_{(+)}|\ell^3}$ 
for $M\ell\approx 0.33$.  
The second case is the one in which $M$ is small compared with $\ell^{-1}$,  
which is the situation discussed in Ref.\cite{GW}.  
In this case, we reproduced the same result found there. 
Namely, the mass 
is approximately given by $(|\varphi_{(-)}|\ell^{3/2}) M^2 \ell$. 
As pointed out in Ref.\cite{GW}, the mass scale on the 
negative tension brane tends to be smaller than the typical 
background energy scale $\ell^{-1}$. We confirmed that 
this is a general feature within the context of weak back 
reaction. 
However, looking at these formulas, 
it seems that we can raise the mass by taking a slightly 
larger vacuum expectation values of the bulk scalar field on both
branes, although a large value of $|\varphi_{(\pm)}|$ is inconsistent
with our approximation. 
Hence, if we remove the technical limitation of weak back reaction, 
it is not clear whether the physical mass of the lowest scalar-type mode 
is always smaller than that of the lowest tensor-type mode. 
We also mention that the mass scale on the
positive tension brane is smaller by a factor of $e^{-d/\ell}$.

Next, we have developed a method to evaluate the explicit form of 
the perturbations caused by the matter fields confined on the branes.  This is
a generalization of the results presented in the previous
paper\cite{GT}.  We found that, as is expected, Einstein gravity is
exactly recovered for the long range force at the order of linear
perturbations when we impose the stabilization mechanism.  The formulas 
for the leading correction from the scalar-type perturbations
(\ref{deltaphiN}) and that from the KK modes (\ref{correctionKK}) are also
derived without specifying the model for the potential of the bulk scalar
field.  From these formulas, we can read the coupling of the metric
perturbations induced on the branes to the matter fields on both branes.

As we have confirmed in Sec.V, there is no pathological behavior  
in all perturbation modes at the level of linear perturbations. 
This is a rather expected result because 
almost all important information at the level of linear perturbations 
is contained in the mass spectrum except for the strength of 
coupling to the matter fields. 
However, it is still unclear what happens once we take into 
account the non-linearity of gravity.  
To investigate this issue, the formulas obtained in this paper 
will be useful. 

\vspace{5mm}
\centerline{\bf Acknowledgements}

We thank J. Garriga for useful comments and discussions. 
T.T. acknowledges support 
from Monbusho System to Send Japanese Researchers Overseas.

\end{multicols}

\begin{thebibliography}{99}
\bibitem{HW} P.~Horava and E.~Witten, Nucl. Phys. {\bf B460}, 506 (1996); 
 Nucl. Phys. {\bf B475}, 94 (1996); 
\bibitem{AH} N.Arkani-Hamed, S.Dimopoulos and G.Dvali, 
Phys. Lett. {\bf B429}, 263 (1998); 
I.~Antoniadis, N.Arkani-Hamed, S.Dimopoulos, G.Dvali, Phys. Lett. 
{\bf B436}, 257 (1998); 
\bibitem{RS1} L.~Randall and R.~Sundrum, Phys. Rev. Lett.{\bf 83}, 
3370 (1999).
\bibitem{RS2} 
L.~Randall and R.~Sundrum, hep-ph/9906064.
\bibitem{Bin} 
P.~Bin\'etruy, C.~Deffayet, U.~Ellwanger and 
D.~Langlois, hep-th/9910219.
\bibitem{Fla}
E.E.~Flanagan, S.-H.H.~Tve and I.~Wasserman, 
hep-ph/9909373. 
\bibitem{CGRT}
C.~Cs\'aki, M.~Graesser, L.~Randall and J.~Terning, 
hep-ph/9911406. 
\bibitem{GS} 
J.~Garriga and M.~Sasaki, hep-th/9912118.
\bibitem{MSM}
S.~Mukohyama, T.~Shiromizu and K.~Maeda, 
hep-th/9912287. 
\bibitem{MWBH}
R.~Maartens, D.~Wands, B.A.~Bassett and I. Heard, 
hep-ph/9912464.
\bibitem{GT}
J.~Garriga and T.~Tanaka, hep-th/9911055 (Paper I).
\bibitem{SMS1} T.~Shiromizu, K.~Maeda and M.~Sasaki, 
gr-qc/9910076.
\bibitem{SMS2} M.~Sasaki, T.~Shiromizu and K.~Maeda, 
gr-qc/9912233.
\bibitem{CG} A.~Chamblin and G.W.~Gibbons, 
hep-th/9909130.
\bibitem{CHR} A.~Chamblin, S.W.~Hawking and H.S.~Reall, 
hep-th/9909205.
\bibitem{EHM} 
R.~Emparan, G.T.~Horowitz and R.C.~Myers, hep-th/9911043.
\bibitem{Chiba}
T.~Chiba, gr-qc/0001029. 
\bibitem{GW}
W.D.~Goldberger and M.B.~Wise, Phys. Rev. Lett.{\bf } (1999), 
hep-ph/9907447;hep-ph/9911457.
\bibitem{DeW}
O.~DeWolfe, D.Z.~Freedman, S.S.~Gubser and A.~Karch, hep-th/9909134.
\bibitem{Rub}
C.~Charmousis, R.~Gregory and V.~Rubakov, 
hep-th/9912160.
\end{thebibliography}
\end{document}